# An Algorithm for Alignment-free Sequence Comparison using Logical Match

Sanil Shanker KP
Department of Computer Science
University of Kerala , India
e-mail: sanilshankerkp@gmail.com

Elizabeth Sherly
Indian Institute of Information
Technology Management- Kerala
India . e-mail: sherly@iiitmk.ac.in

Jim Austin
Department of Computer Science
University of York, UK
e-mail: austin@cs.york.ac.uk

**Abstract:** This paper proposes an algorithm for alignment-free sequence comparison using Logical Match. Here, we compute the score using fuzzy membership values which generate automatically from the number of matches and mismatches. We demonstrate the method with both the artificial and real datum. The results show the uniqueness of the proposed method by analyzing DNA sequences taken from NCBI databank with a novel computational time.

## I   INTRODUCTION

Alignment-free sequence comparison remains a computational problem as the total number of sequences in the underlying databases grows exponentially with the progress of research[1]. Algorithms devised for the comparison of molecular sequences are based on the concept of string matching[2]. The newly proposed alignment-free sequence comparison algorithm is based on the concept of string matching. techniques. In Brute Force method, the string matching algorithm compares a pattern character by character in each and every location of the text. It is possible to solve the problems of strings matching with the help of finite automata[3]. Using finite automata, string matching automation is built from the pattern as a preprocessing step before matching. The text is then scanned through the automation to find occurrences of the pattern in the text. The Knuth-Morris-Pratt algorithm avoids back-tracking on the text when a mismatch occurs, by exploiting the knowledge of the matched substring in the text prior to the mismatch[4]. The main peculiarity of the Boyer-Moore algorithm is that some of the characters in the text can be skipped completely without comparing them with the pattern as it can be shown that they can never contribute to an occurrence of the pattern in the text[5]. In Logical Match, the sequence is arranged so that each element in the pattern are encoded as binary digits and coincides with it's corresponding index  and then proceed to match logically the indices of the pattern with those of the text. This paper presents the method for alignment-free comparison of sequential patterns of finite length using automatically generating fuzzy membership values by Logical Match[6,7,8,9].

## II   METHOD

The sequence pattern is arranged so that each characters are encoded as binary digits and coincide with the corresponding index and then proceed to match logically the indices of the pattern with those of the text. The algorithm has two phases. The characters in the sequence pattern are pre- processed in the phase- I. The information from the phase- 1 is used in the phase- II in order to reduce the total number of character comparisons. We compute the score using fuzzy membership values which generate automatically  from the number of matches and mismatches.

Phase - I

i, Each characters are encoded as binary digits and generate the indices of Text and Pattern using Logical Match

Phase - II

ii, Compute number of Match(Text,Pattern) and Mismatch(Text,Pattern)

iii, Compute score,  S(Text,Pattrn)

$= $ Match in Text $*\mu_{Match(Pattern)}[Pattern] - $

Mismatch in Text$*\mu_{Mismatch(Pattern)}[Pattern]$ ,

where, $\mu_{Match(Pattern)}[Pattern] + \mu_{Mismatch(Pattern)}[Pattern] = 1$

( See Appendix )

## III SIMULATION WITH ARTIFICIAL DATUM

Text => <ATCAAGATCA>
Pattern => <AAGAGGCTCA>

a) Phase- I: Generating indices of Text and Pattern using Logical Match

Initialize {1000←A ,0100←T,0010←G,0001←C}
Shift the text(Table 1) and pattern(Table 2) so that each encoded binary digit in the sequence coincides with it's corresponding index in it's respective column.

Text => < 1000(1,4,6,7,10);
0100(3,9);
0010(5);
0001(2,8) >

|    | 0/1 | 0/1 | 0/1 | 0/1 |
|----|-----|-----|-----|-----|
| 10 | 1   | 0   | 0   | 0   |
| 9  | 0   | 1   | 0   | 0   |
| 8  | 0   | 0   | 0   | 1   |
| 7  | 1   | 0   | 0   | 0   |
| 6  | 1   | 0   | 0   | 0   |
| 5  | 0   | 0   | 1   | 0   |
| 4  | 1   | 0   | 0   | 0   |
| 3  | 0   | 1   | 0   | 0   |
| 2  | 0   | 0   | 0   | 1   |
| 1  | 1   | 0   | 0   | 0   |

Table 1. Text(Phase- 1)

Pattern => < 1000(1,7,9,10);
0100(3);
0010(5,6,8);
0001(2,4) >

|    | 0/1 | 0/1 | 0/1 | 0/1 |
|----|-----|-----|-----|-----|
| 10 | 1   | 0   | 0   | 0   |
| 9  | 0   | 0   | 0   | 1   |
| 8  | 0   | 0   | 1   | 0   |
| 7  | 1   | 0   | 0   | 1   |
| 6  | 0   | 0   | 1   | 0   |
| 5  | 0   | 0   | 1   | 0   |
| 4  | 0   | 0   | 0   | 1   |
| 3  | 0   | 1   | 0   | 0   |
| 2  | 0   | 0   | 0   | 1   |
| 1  | 1   | 0   | 0   | 0   |

Table 2. Pattern(Phase- 1)

b) Phase-II:
Compute the number of match and mismatch using Logical Match(Table 3).

| Encoded binary digits | Indices | | | | Match/ Mismatch |
|---|---|---|---|---|---|
| 1000 | **1** | 7 | 9 | 10 | 1st Match |
| 0001 | **2** | 4 |   |   | 2nd Match |
| 0100 | **3** |   |   |   | 3rd Match |
| 1000 | 1 | 7 | 9 | 10 | 1st Mismatch |
| 0010 | **5** | 6 | 8 |   | 4th Match |
| 1000 | 1 | 7 | 9 | 10 | 2nd Mismatch |
| 1000 | 1 | **7** | 9 | 10 | 5th Match |
| 0001 | 2 | 4 |   |   | 3rd Mismatch |
| 0100 | 3 |   |   |   | 4th Mismatch |
| 1000 | 1 | 7 | 9 | **10** | 6th Match |

Table 3. Alignment-free comparison using Logical Match

Compute the score using automatically generating membership values.

Score =

= Match in Text * $\mu_{Match(Pattern)}$[Pattern] −

Mismatch in Text * $\mu_{Mismatch(Pattern)}$[Pattern]

= 6* 0.6 - 4* 0.4 = 3.6-1.6 = 2

## IV EXPERIMENTAL RESULTS

To simulate, alignment–free sequence comparison using Logical Match, the program has been written in C++ language under Linux platform. The method was tested against DNA sequences, the inputs have taken from the Locus ACU90045 as common text and ACU90045, PAU90054, HSU90049, LPU90051, NAU90053, DCU90047,

DPU90048 as patterns of common range 541-560 from NCBI databank(Table 4). In the phase- I of the algorithm, the time complexity is O(m+n) and in the phase- II, the computational time depends on the lengths of the elements in the pattern of the text.

## V CONCLUSION

We have presented the algorithm for alignment-free sequence comparison using Logical Match. The method provides a solution to find alignment- free similarities between two finite sequences by calculating the score using automatically generating fuzzy membership values. This procedure can possibly be implemented in the applications related to the alignment-free comparison of sequential patterns.

Table 4

| Locus | Region: 541-560(20bp) | Score | Match(%) |
|---|---|---|---|
| T: ACU90045 | cgacctctggacaggccact | | |
| P ACU90045 | cgacctctggacaggccact | 20 | 100% |
| T: ACU90045 | cgacctctggacaggccact | | |
| P: PAU90054 | cgacccactgagaaacctct | -2 | 45% |
| T: ACU90045 | cgacctctggacaggccact | | |
| P: HSU90049 | cgaccaactgacaaggctct | -6 | 35% |
| T: ACU90045 | cgacctctggacaggccact | | |
| P: LPU90051 | cgtcccactgacaagcctct | -8 | 30% |
| T: ACU90045 | cgacctctggacaggccact | | |
| P: NAU90053 | cgcccaactgacaaggctct | -10 | 25% |
| T: ACU90045 | cgacctctggacaggccact | | |
| P: DCU90047 | aggcctttggacaaacctct | -12 | 20% |
| T: ACU90045 | cgacctctggacaggccact | | |
| P: DPU90048 | agaccagttgacaaaccttt | -16 | 10% |

Where, T and P are Text and Pattern respectively.

## ACKNOWLEDGMENT

SSKP was funded in part by European Research and Educational Collaboration with Asia

## APPENDIX

Let T and P be two strings of lengths n and m respectively, where n ≥ m. When P compares(alignment-free) with T gives r matches and s mismatches, r + s = m

$\mu_{Match(P)}[P] + \mu_{Mismatch(P)}[P]$ = Match in P/|P| + Mismatch in P/|P|

= (Match in P + Mismatch in P)/|P|

= (r + s) / m = m / m = 1

Example **a**)

```
T  => 0 1 0 1 0 0 1 0
      | | | # # | # |
P  => 0 1 0 0 1 0 0 0
```

$\mu_{Match(Pattern)}[P] = 5/8 = 0.625$,
$\mu_{Mismatch(Pattern)}[P] = 3/8 = 0.375$

Score(T ,P) = Match in T *$\mu_{Match(P)}[P]$ −

Mismatch in T*$\mu_{Mismatch(P)}[P]$

$= 5*0.625 - 3* 0.375 = 3.125 - 1.125 = 2$

Example **b**)

```
T  => 0 1 0  1 0 0 1 0
        | | |  # # | # #
P  => 0 1 0 0 1 0 0 --
```

$\mu_{Match(Pattern)}[P] = 4/7 = 0.571$
$\mu_{Mismatch(Pattern)}[P] = 3/7 = 0.428$

Score(T ,P) = Match in T * $\mu_{Match(P)}[P]$ −

Mismatch in T * $\mu_{Mismatch(P)}[P]$
$= 4*0.571 - 4* 0.428 = 2.284 - 1.712$
$= 0.572$

Example **c**)

```
T  => 0 1 0  1 0 0 1 0
        | | |  # # # # #
P  => 0 1 0 0 -- -- -- --
```

$\mu_{Match(Pattern)}[P] = 3/4 = 0.75$,
$\mu_{Mismatch(Pattern)}[P] = 1/4 = 0.25$

Score(T ,P) = Match in T * $\mu_{Match(P)}[P]$ −

Mismatch in T * $\mu_{Mismatch(P)}[P]$
$= 3*0.75 - 5* 0.25 = 2.25 - 1.25$
$= 1$